\documentclass{aa} 

\input epsf

\usepackage{graphicx}

\usepackage[mathcal]{eucal}
\usepackage{amssymb}
\usepackage{amsmath}
\usepackage{natbib}
\usepackage[T1]{fontenc}
\def\ds {$\delta$\,Scuti}
\def\gd {$\gamma$\,Doradus}

\def\cd {d$^{-1}$}

\begin{document}


\title{The CoRoT star ID\,100866999: a hybrid \gd-\ds\ star in an eclipsing binary system.
\thanks{The CoRoT space mission was developed and is operated by the French 
space agency CNES, with participation of ESA's RSSD and Science Programmes, 
Austria, Belgium, Brazil, Germany, and Spain.}}

\author{E. Chapellier \inst{1}, P. Mathias \inst{2,} \inst{3}
}  

\institute{
Laboratoire Lagrange, UMR7293, Universit\'e de Nice Sophia-Antipolis, CNRS,
Observatoire de la C\^ote d'Azur, 06300 Nice, France
\and
Universit\'e de Toulouse; UPS-OMP; IRAP;  F-65000 Tarbes, France
\and
CNRS; IRAP; 57, Avenue d'Azereix, BP 826, F-65008 Tarbes, France
}
\date{Received date; accepted date}

\authorrunning{E. Chapellier and P. Mathias}
\titlerunning{Couplings in an eclipsing hybrid \gd-\ds\ star}


 
\abstract
{The presence of $g$- and $p$-modes allows testing stellar models from the core to the envelope. 
Moreover, binarity in an eclipsing system constrains the physical parameters of the pulsating star.
}
{CoRot\,ID\,100866999 is a relatively large-amplitude hybrid \gd-\ds\ star with two clearly distinct 
frequency domains. 
The large number of detected frequencies allows a detailed study of the interaction between them. 
In addition, we can derive the fundamental parameters of both components from the study of the eclipsing light curve. 
} 
{After removing the eclipsing phases, we analyzed the data with the Period04 package up to a signal-to-noise ratio 
S/N=4. 
The light curve was then prewhitened with these oscillation frequencies to derive the fundamental 
parameters of the two components. 
}
{The eclipsing light curve analysis results in a (1.8+1.1)\,M$_{\odot}$ system, both components being
main sequence stars.
We detect 124 frequencies related to luminosity variations of the primary. 
They are present in two well-separated domains: 89 frequencies in the interval 
[0.30;3.64]\,\cd\ and 35 in the interval [14.57; 33.96]\,\cd. 
There are 22 \gd\ frequencies separated by a constant period interval $\Delta P = 0.03493$\,d. 
These frequencies correspond to a series of g-modes of degree $\ell=1$ with successive 
radial orders $k$. 
We identify 21 linear combinations between the first nine \gd\ frequencies. 
The \ds\ domain is dominated by a large-amplitude frequency $F=16.9803$\,\cd. 
The eight first \gd\ frequencies $f_i$ are present with much lower amplitude in the \ds\ domain as 
$F\pm f_i$. 
These interactions between $g$- and $p$-modes confirm the phenomenon we detected in another CoRoT star. \\
The amplitude and the phase of the main frequency $F$ shows a double-wave modulation along the orbital 
phase, giving rise to series of combination frequencies. 
Such combination frequencies are also detected, with lower amplitude, for the first \gd\ modes.
}
{}


\maketitle

\section{Introduction}

The launch of several dedicated satellites (MOST \citep{wmk03}, CoRoT \citep{betal06}, Kepler \citep{bkd97,cab08}),
which allow the detection of several hundreds of accurate frequencies, opened a new era in asteroseismic studies.
Concerning the classical instability strip, preliminary results show, for instance that the regions of \ds\ and \gd\ 
stars largely overlap \citep{umg11}, with the presence of low frequencies in a large fraction of A-F stars
\citep{b11}.
However, from a theoretical point of view, models fail to reproduce clearly such hybrid stars, because the two identified mechanisms
($\kappa$-mechanism for high frequencies, convective blocking mechanism for low frequencies) work together on only 
a small fraction of the observed instability strip.

In a preceding paper (\citet{cmw12}, hereafter Paper\,I), we detected strong coupling between $g$- and $p$-modes in a 
hybrid \gd-\ds\ star: CoRoT\,ID\,105733033 (hereafter 033).  
Each $g$-mode frequency $f_i$ gives rise to two smaller amplitude frequencies $F-f_i$ and $F+f_i$ around the main $g$-mode 
frequency $F$. 
Such a phenomenon has been predicted for the Sun \citep{kjh93} and interpreted as a coupling between the $g$- and $p$-modes cavities. 
If valid, this physical interpretation should act in other hybrid \gd-\ds\ stars.
We thus chose a candidate with similar pulsation characteristics, a hybrid with a dominant 
\ds\ mode and several \gd\ modes. 
To obtain a precise value of the rotation period, we selected an eclipsing binary star, CoRoT\,ID\,100866999 (hereafter 999). 
The data are described in Sect.\,2, and the eclipsing binary system is studied in Sect.\,3.
The frequency analysis is presented in Sect.\,4, and results are studied in Sect.\,5 and 6 for the low- and high-frequency 
regions, respectively.
Section\,7 deals with the relationship between the different pulsation frequencies and the orbital one.
Finally, conclusions are discussed in Sect.\,8.

\section{The CoRoT data}

\begin{figure}
\centering
\includegraphics[width=8cm]{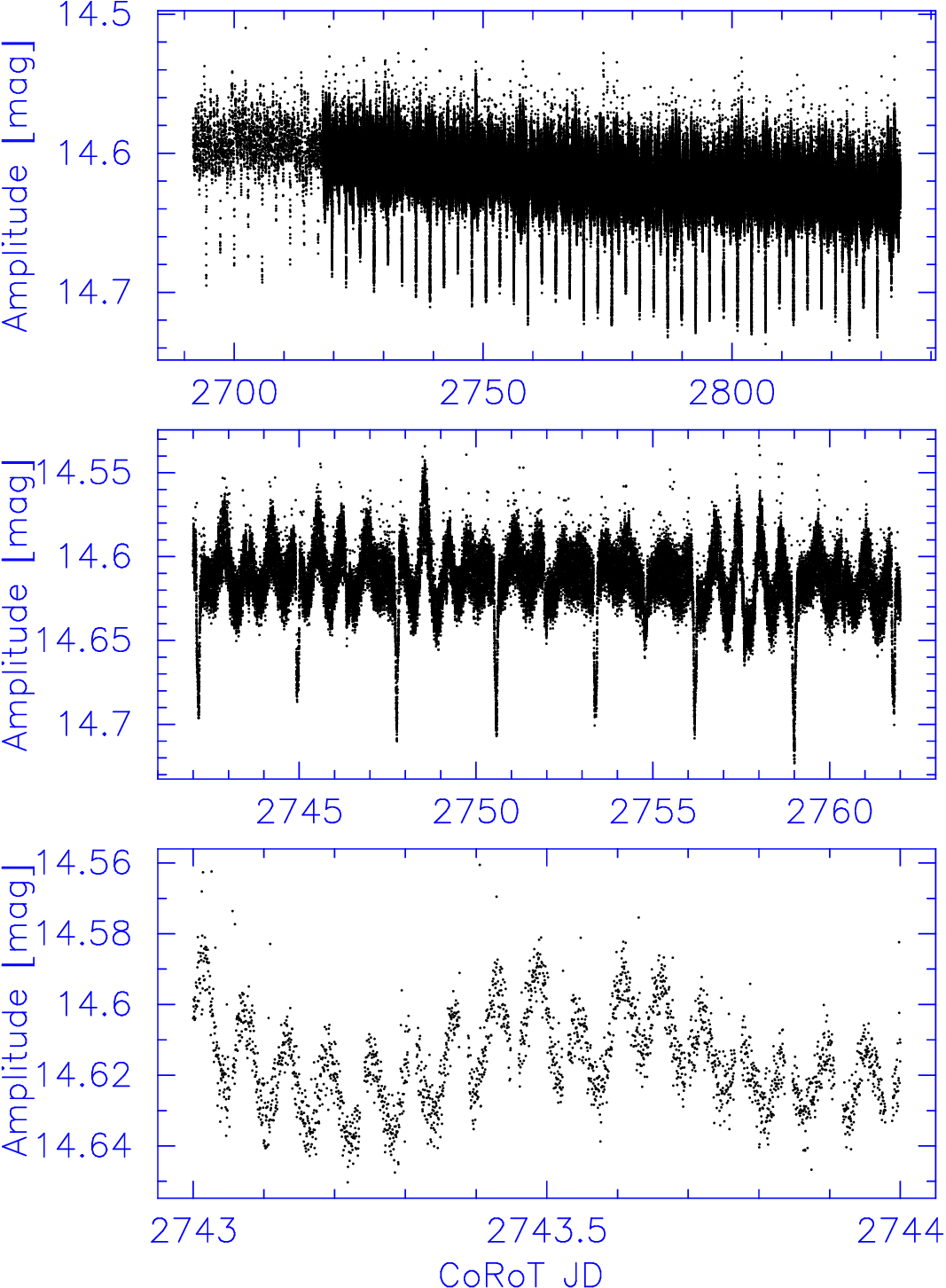}
\caption[]{Light curve of the binary system 999 with different timescales.
From top to bottom, the complete light curve over 142\,d, then a subset over 22\,d and finally a zoom into a
3.5\,d subset. Eclipse phases are well determined.
We note the different sampling: 512\,s and 32\,s.}
\label{fig01}
\end{figure}

The observations of 999 were collected during CoRoT's first long run,
LRc01, which targeted the Galactic center.
We worked on the reduced N2 light curves \citep{abb09} throughout this paper.
The observations lasted 142 days, from May 16$^{\rm th}$ to October 5$^{\rm th}$, 2007.
Among the 317045 measurements obtained, we retained only the data flagged ``0'' by the
CoRoT pipeline because the other measurements were affected by instrumental effects such as straylight, 
cosmic rays, and perturbation by Earth eclipses.
In the following, the measured flux was converted into magnitude using the CoRoT magnitude $C=14.599$.
This value actually corresponds to the $R$-magnitude provided by the EXODAT database \citep{dmm09}.
The resulting light curve, on different timescales, is represented in Fig.\,\ref{fig01}.
The light curve is recorded in "white" light, i.e., no color information is available.
The measurements were obtained at different sampling (512\,s during the first 26 days and then 
32\,s during the other 116 days).
For the coherence of the data, all the measurements were binned to a common sampling of 512\,s. 
In Fig.\,\ref{fig01}, the presence of low and high frequencies is obvious.
In the following, the timescale is labeled in CoRoT Julian Day, where the starting CoRoT
JD corresponds to HJD\,2451545.0 (January 1$^{\rm st}$, 2000 at UT 12:00:00).

CoRoT data are known to be affected by several technical issues, such as long-term trends
and jumps due to cosmic rays \citep{abb09}.
In addition, many individual measurements can be considered as outliers. 
The most significant of them 
were removed by an iterative procedure during the Fourier analysis.
To correct the trends and the jumps, we use a new and original procedure. 
We first performed a Fourier analysis of the binned data up to 200 frequencies and then prewhitened 
the data with all the frequencies higher than 0.25\,\cd. 
We applied a spline function in the residuals and then prewhitened the original data set with the same spline function. 
This procedure gave a satisfying straightening of the data.

The EXODAT database \citep{dmm09} provides two possible spectral types: K5\,V or A5\,IV and a color temperature 
(4900\,K) that is compatible with none of these spectral identifications. 
The magnitudes from the 2MASS survey ($J=12.95$, $H=12.59$, $K=12.40$) are more compatible with an A5\,IV spectrum. 
From a survey of CoRoT stars candidates using low-resolution spectroscopy, \citet{sdn13} obtained an effective
temperature of the order of $7700 \pm 400$\,K, which is not compatible with the K5 type.
Because the presence of a companion perturbs spectral type identification, we must consider that we have no 
precise parameters for this star. 
However we can exclude a K5\,V spectrum that puts the star outside of the well-determined \ds\ instability strip.

The CoRoT contamination factor is very small (4.0\,\%), so we consider the signal as coming
from the eclipsing binary system of 999 alone.

\section{The eclipsing binary system}

An exploration of the LRc01 field for transit detection by \citet{cfo09} showed that the
star is an eclipsing binary (E2-1719) with an orbital period of $2.808769 \pm 0.000022$\,d and an epoch
of $2\,454\,239.396628 \pm 0.000656$\,HJD. 
The eclipse lasts 3.9\,h with a depth of 6.7\,\%.
Two eclipses are visible, separated by half the orbital period, implying a null eccentricity.
The duration of each eclipse is equal to 5.3\,\% of the orbital period. 

To determine the parameters of both components, we used the PHOEBE software \citep{pz05}.
From the Fourier analysis, we fixed both the orbital period and the null eccentricity.
The light curve was first filtered from all the frequencies that cannot be attributed to the binary motion,
in particular, the pulsation frequencies, but also the satellite ones.
We note that the CoRoT data in white light are the only data at our disposal, so 
our physical determinations are relatively imprecise due to the lack of color light curves 
and radial velocities.

\begin{figure}
\centering
\includegraphics[width=6.5cm,angle=-90.]{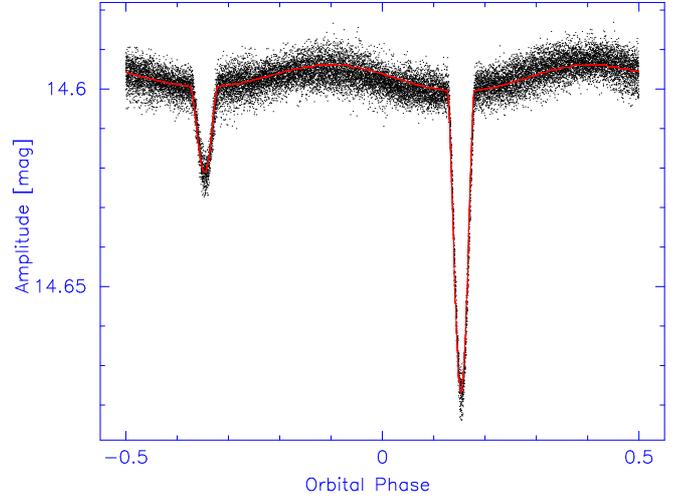}
\caption[]{Light curve of the binary system 999 after prewhitening with the 
pulsation frequencies. 
The red line represents the fit obtained with the orbital parameters given in Table\,1.}
\label{fig02}
\end{figure}

Several attempts led to the determination of the orbital parameters listed in Table\,1. 
\begin{table}
\begin{center}
\caption[]{Resulting parameters from the eclipsing curve fit.}
\begin{tabular}{lcc}
\hline
\multicolumn{1}{c}{Parameter} &
\multicolumn{2}{c}{System} \\
\hline
Period [d] & \multicolumn{2}{c}{$2.80889 \pm 0.00011$} \\
Inclination [$^{\circ}$] & \multicolumn{2}{c}{$80 \pm 2$} \\
a [$R_{\odot}$] & \multicolumn{2}{c}{$12 \pm 0.8$} \\
q [M$_2$/M$_1$] & \multicolumn{2}{c}{$0.65 \pm 0.07$} \\

\hline
\multicolumn{1}{c}{} &
\multicolumn{1}{c}{Primary} &
\multicolumn{1}{c}{Secondary} \\
\hline
Mass   $[M_{\odot}]$ & $1.8 \pm 0.2$ & $1.1 \pm 0.2$ \\
Radius $[R_{\odot}]$ & $1.9 \pm 0.2$ & $0.9 \pm 0.2$ \\
Log g $[{\rm cgs}]$ & $4.1 \pm 0.1$ & $4.6 \pm 0.1$ \\
T$_{\rm eff}$ $[K]$  & $7300 \pm 250$ & $5400 \pm 430$ \\
M$_{\rm bol}$ & $2.3 \pm 0.2$ & $5.4 \pm 0.3$ \\
\hline
\end{tabular}
\end{center}
\label{tab1}
\end{table}
The related fit is represented in Fig.\,\ref{fig02}.
From these results, it appears that the primary star is compatible with an A7-F0 spectral type,
while the secondary star is compatible with a G5-KO spectral type, both being on the main sequence.
Therefore, the primary star is within the ``classical'' instability strip, while the secondary one is not.

To be sure that the pulsation really originates from the primary star, we analyzed separately the data obtained only  
during the secondary eclipse (when the companion is hidden by the primary). 
The detection of the first \ds\ and \gd\ frequencies in the Fourier analysis i.e., those having the largest amplitudes, 
ensures that the primary is the pulsating star.

\section{Frequency analysis}

\begin{figure}
\centering
\includegraphics[width=8cm]{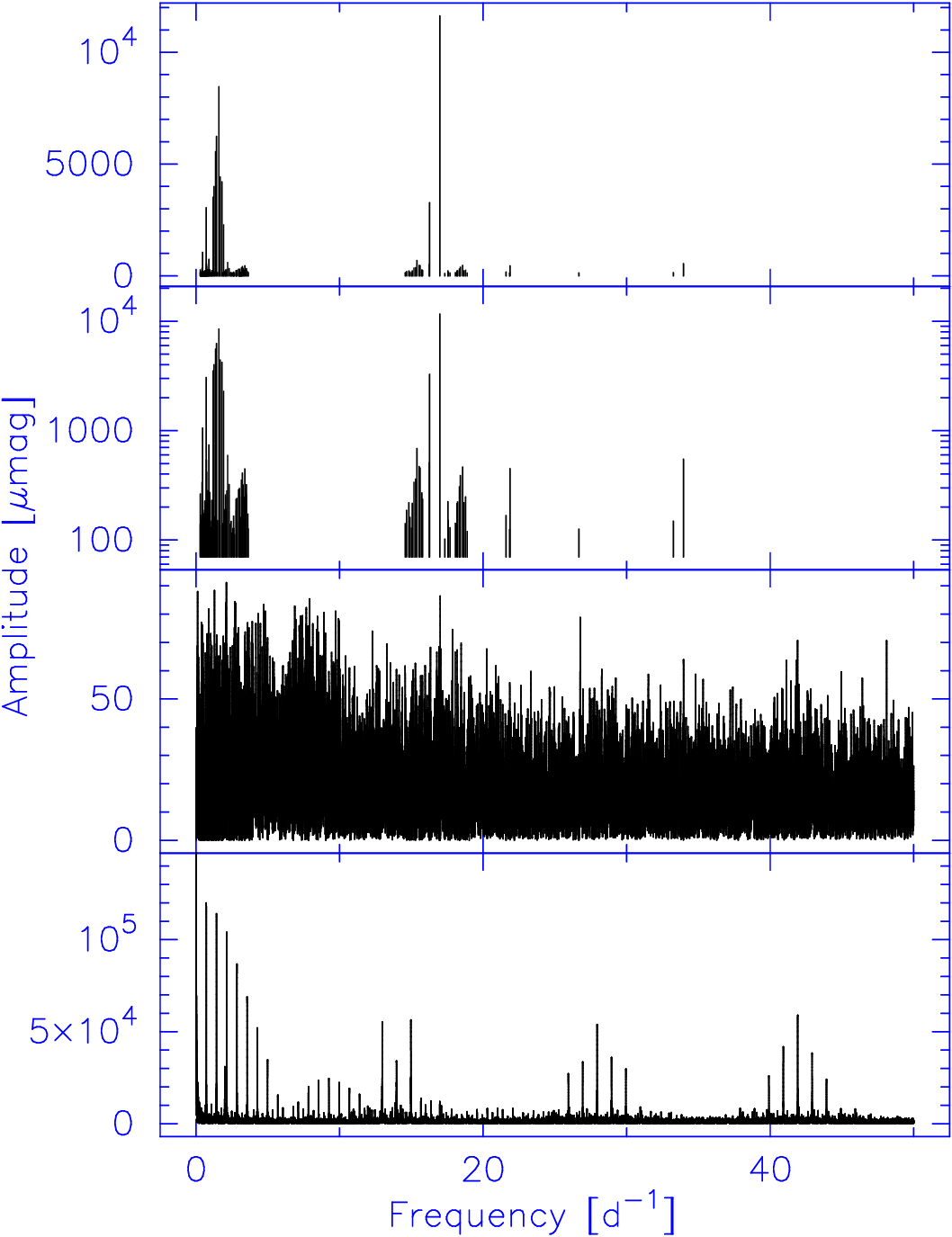}
\caption[]{Distribution of the 124 retained frequencies.
From top to bottom: the amplitudes in
$\mu$mag, and the same with a logarithmic scale,
the Fourier spectrum computed from the residuals,
and the spectral window.}
\label{fig03}
\end{figure}

The data were analyzed with the package Period04 \citep{lb05}. 
To detect only the pulsation frequencies, we removed the data obtained during both eclipses.
After re-sampling and elimination of the outliers, we ended with 19207 measurements.
The frequencies were searched in the interval $[0;84]$\,\cd\ i.e., up to the Nyquist limit. 
For each detected frequency, the amplitude and the phase were calculated by a least-squares sine fit. 
The data were then prewhitened and a new analysis was performed on the residuals. 
The Fourier analysis was stopped after 200 frequencies, when we reached the usual S/N=4 limit, even in the 
less noisy spectral regions.
We note that we did not use the option {\tt improve all} available in the Period04 package, which computes a multi-sine fit.
Indeed, we noted that when the number of frequencies increases, the narrowest couple of frequencies tend to converge to
an unique one. 
To compensate for the loss of precision when this option is not considered, we used a very high step rate in the Fourier 
analysis (0.0001\,\cd). 
The uncertainties in the frequencies, amplitudes, and phases were computed with the formulae proposed by \citet{mo99}.
The final noise varies from 0.031\,mmag in the \gd\ region to 0.024\,mmag in the \ds\ region. 
We eliminated 47 frequencies having an S/N value lower than four. 
We also excluded the frequencies lower than 0.25\,\cd. 
When two frequencies were closer than $1/\Delta T=0.007$\,\cd, we systematically discarded the lowest amplitude frequency. 
After this elimination process, 124 frequencies were retained: 89 in the interval $[0.30;3.64]$\,\cd\ and 35 in the interval 
$[14.57;33.96]$\,\cd\ (Table\,2, Fig.\,\ref{fig03}).
\begin{table*}
\begin{center}
\caption[]{The first 25 stellar frequencies (the complete table is available
electronically).}
\begin{tabular}{lrrrrrrcr}
\hline
\multicolumn{2}{c}{Frequency} &
\multicolumn{1}{c}{$\sigma$} &
\multicolumn{1}{c}{A} &
\multicolumn{1}{c}{$\Phi$} &
\multicolumn{1}{c}{$\sigma$} &
\multicolumn{1}{c}{S/N} &
\multicolumn{1}{c}{Ident.} &
\multicolumn{1}{c}{$k$} \\
\multicolumn{1}{c}{} &
\multicolumn{1}{c}{\cd} &
\multicolumn{1}{c}{\cd} &
\multicolumn{1}{c}{mmag} &
\multicolumn{1}{c}{rad} &
\multicolumn{1}{c}{rad} &
\multicolumn{1}{c}{ } &
\multicolumn{1}{c}{} &
\multicolumn{1}{c}{} \\
\hline
$F_1$    & 16.98030 & 0.00001 & 11.623 & 3.462 & 0.001 & 515.5 & $F$              &    \\
$F_2$    &  1.59540 & 0.00001 &  8.455 & 3.914 & 0.002 & 301.4 & $f_1$            &  0 \\
$F_3$    &  1.43420 & 0.00001 &  6.237 & 0.151 & 0.003 & 225.0 & $f_2$            &  2 \\
$F_4$    &  1.36600 & 0.00001 &  5.551 & 1.307 & 0.003 & 201.3 & $f_3$            &  3 \\
$F_5$    &  1.69030 & 0.00001 &  4.417 & 5.529 & 0.004 & 156.3 & $f_4$            & -1 \\
$F_6$    &  1.79740 & 0.00002 &  4.206 & 3.129 & 0.004 & 147.6 & $f_5$            & -2 \\
$F_7$    &  1.24880 & 0.00002 &  3.993 & 2.073 & 0.004 & 146.0 & $f_6$            &  5 \\
$F_8$    &  1.19840 & 0.00002 &  3.517 & 5.385 & 0.005 & 129.0 & $f_7$            &  6 \\
$F_9$    & 16.26820 & 0.00002 &  3.271 & 5.246 & 0.005 & 146.5 & $F-2f_{\rm orb}$ &    \\
$F_{10}$ &  0.71190 & 0.00002 &  3.051 & 0.930 & 0.005 & 114.8 & $2f_{\rm orb}$   &    \\
$F_{11}$ &  1.91650 & 0.00003 &  2.278 & 0.038 & 0.007 &  79.2 & $f_8$            & -3 \\
$F_{13}$ &  0.45150 & 0.00006 &  1.054 & 4.279 & 0.016 &  39.9 & $f_9$            &    \\
$F_{15}$ &  0.89730 & 0.00009 &  0.738 & 3.380 & 0.023 &  27.5 & $f_{10}$         & 14 \\
$F_{16}$ & 15.38510 & 0.00009 &  0.684 & 0.383 & 0.024 &  31.6 & $F-f_1$          &    \\
$F_{17}$ &  2.20150 & 0.00011 &  0.591 & 2.249 & 0.028 &  20.1 & $f_{11}$         & -5 \\
$F_{18}$ & 33.96040 & 0.00012 &  0.547 & 0.214 & 0.030 &  29.0 & $2F$             &    \\
$F_{19}$ &  0.72790 & 0.00012 &  0.535 & 5.510 & 0.031 &  20.1 & $f_{12}$         &    \\
$F_{21}$ & 16.25300 & 0.00013 &  0.508 & 0.440 & 0.033 &  22.8 & $p_1$            &    \\
$F_{22}$ & 15.54600 & 0.00014 &  0.468 & 3.072 & 0.036 &  21.5 & $F-f_2$          &    \\
$F_{23}$ & 18.57580 & 0.00014 &  0.464 & 0.641 & 0.036 &  22.5 & $F+f_1$          &    \\
$F_{24}$ & 15.61460 & 0.00014 &  0.453 & 1.225 & 0.037 &  20.8 & $F-f_3$          &    \\
$F_{25}$ & 21.87110 & 0.00014 &  0.449 & 4.486 & 0.037 &  23.3 & $p_2$            &    \\
$F_{26}$ &  3.39280 & 0.00014 &  0.446 & 2.639 & 0.037 &  14.7 & $f_1+f_5$        &    \\
$F_{28}$ &  0.75010 & 0.00016 &  0.415 & 0.949 & 0.040 &  15.6 & $f_{13}$         &    \\
$F_{30}$ &  3.23170 & 0.00016 &  0.409 & 3.556 & 0.041 &  13.4 & $f_2+f_5$        &    \\
\hline
\end{tabular}
\tablefoot{
Successive columns respectively correspond to the 
frequency detection order, its value together with its associated sigma, the corresponding amplitude
(the related sigma value is 0.017\,mmag), the phase and its sigma, and S/N.
We also indicate the identification and the eventual $k$ radial order
(relative to $f_1$) of the asymptotic $g$-modes.}
\end{center}
\label{tab2}
\end{table*}

Removing the data obtained during the eclipses leads to the large pike aliases 
shown in Fig.\,\ref{fig03}. 
To detect eventual alias effects on the Fourier analysis, we undertook a second 
analysis on the entire data set after prewhitening it from the binary model. 
We performed the same Fourier analysis up to 200 frequencies. 
In this analysis, only the aliases related to the satellite frequency remained, but the noise 
increased by 10 to 30\,\% in the \gd\ region.
No significant differences are observed between the frequencies detected with the two methods. 
In particular, none of the 124 frequencies seems to be affected by aliasing effects. 
As the first method gave the best S/N, we only consider these results.

The \gd\ and the \ds\ domains are well separated; no frequencies are detected in the interval $[3.64;14.57]$\,\cd.

\section{The \gd\ domain}

\begin{figure}
\centering
\includegraphics[width=6.5cm,angle=-90.]{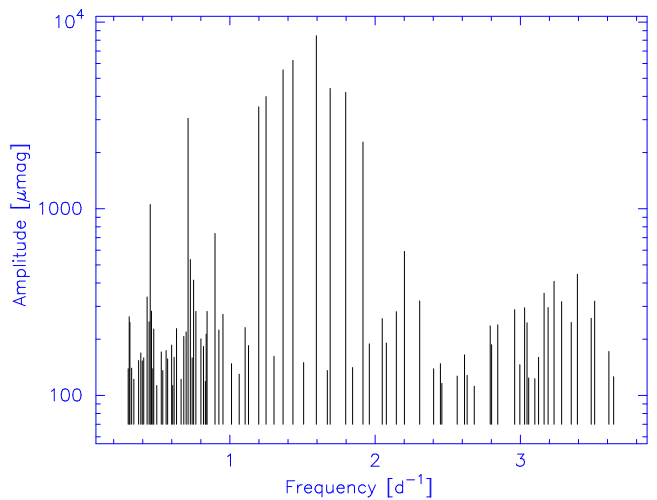}
\caption[]{Detected frequencies in the \gd\ domain.
The ordinates are on a logarithmic scale.
}
\label{fig04}
\end{figure}

\begin{figure}
\centering
\includegraphics[width=6.5cm,angle=-90.]{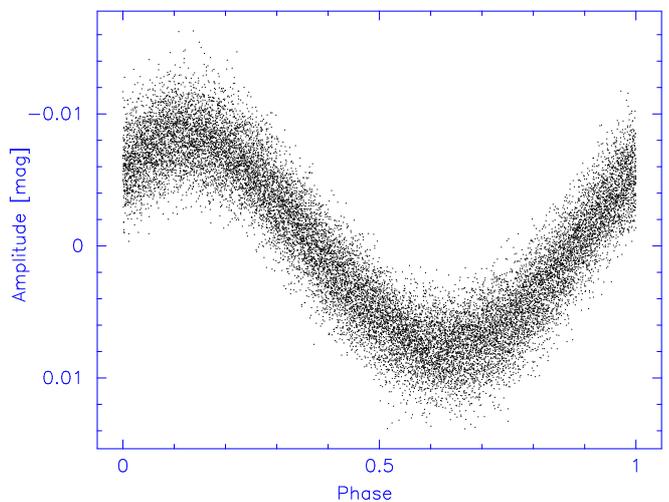}
\caption[]{Data phased with the main \gd\ frequency $f_1=1.5954$\,\cd.
}
\label{fig05}
\end{figure}

From the Fourier analysis, we detected 89 frequencies in the interval $[0.30;3.64]$\,\cd\ (Table\,2, Fig.\,\ref{fig04}). 
The main frequency is $f_1=1.5954$\,\cd, with an amplitude $A_1=8.45$\,mmag. 
Data phased with this frequency are represented in Fig.\,\ref{fig05}. 
The shape of the curve is nearly sinusoidal, hence only $2f_1$ is detected with an amplitude $A=0.3$\,mmag. 
The light variations associated with the ellipsoidal deformation of the main star are detected as $2 f_{\rm orb}=0.7119$\,\cd\ 
with an amplitude $A=3.12$\,mmag. 
We identified 21 linear combinations between the first nine \gd\ frequencies (Table\,3). 
\begin{table}
\begin{center}
\caption[]{Linear combinations of frequencies detected in the \gd\ domain.}
\begin{tabular}{lcr}
\hline
\multicolumn{1}{c}{Frequency} &
\multicolumn{1}{c}{Combination} &
\multicolumn{1}{c}{$\Delta$} \\
\hline
$F_{45}$  & $f_1+f_2$ &  6 \\
$F_{46}$  & $f_1+f_3$ &  9 \\
$F_{40}$  & $f_1+f_4$ & 15 \\
$F_{26}$  & $f_1+f_5$ &  0 \\
$F_{59}$  & $f_1+f_6$ &  0 \\
$F_{63}$  & $f_1+f_7$ & 16 \\
$F_{41}$  & $f_1+f_8$ &  6 \\
$F_{60}$  & $f_1+f_9$ & 21 \\
$F_{95}$  & $f_2+f_3$ & 18 \\
$F_{117}$ & $f_2+f_4$ &  9 \\
$F_{30}$  & $f_2+f_5$ &  1 \\
$F_{161}$ & $f_2+f_6$ &  6 \\
$F_{125}$ & $f_2+f_7$ &  7 \\
$F_{65}$  & $f_2+f_8$ &  7 \\
$F_{152}$ & $f_3+f_4$ &  8 \\
$F_{33}$  & $f_3+f_5$ &  1 \\
$F_{109}$ & $f_3+f_6$ &  6 \\
$F_{142}$ & $f_3+f_7$ &  7 \\
$F_{53}$  & $f_4+f_5$ &  1 \\
$F_{58}$  & $f_5+f_6$ &  2 \\
$F_{119}$ & $f_5+f_7$ &  7 \\
\hline
\end{tabular}
\tablefoot{
Columns are respectively the label of the frequency,
the linear combination, and the differences $\Delta$ between the linear combination and the actually measured
values [$\times 10^{-4}$\,\cd].}
\end{center}
\label{tab3}
\end{table}

\begin{figure}
\centering
\includegraphics[width=6.5cm]{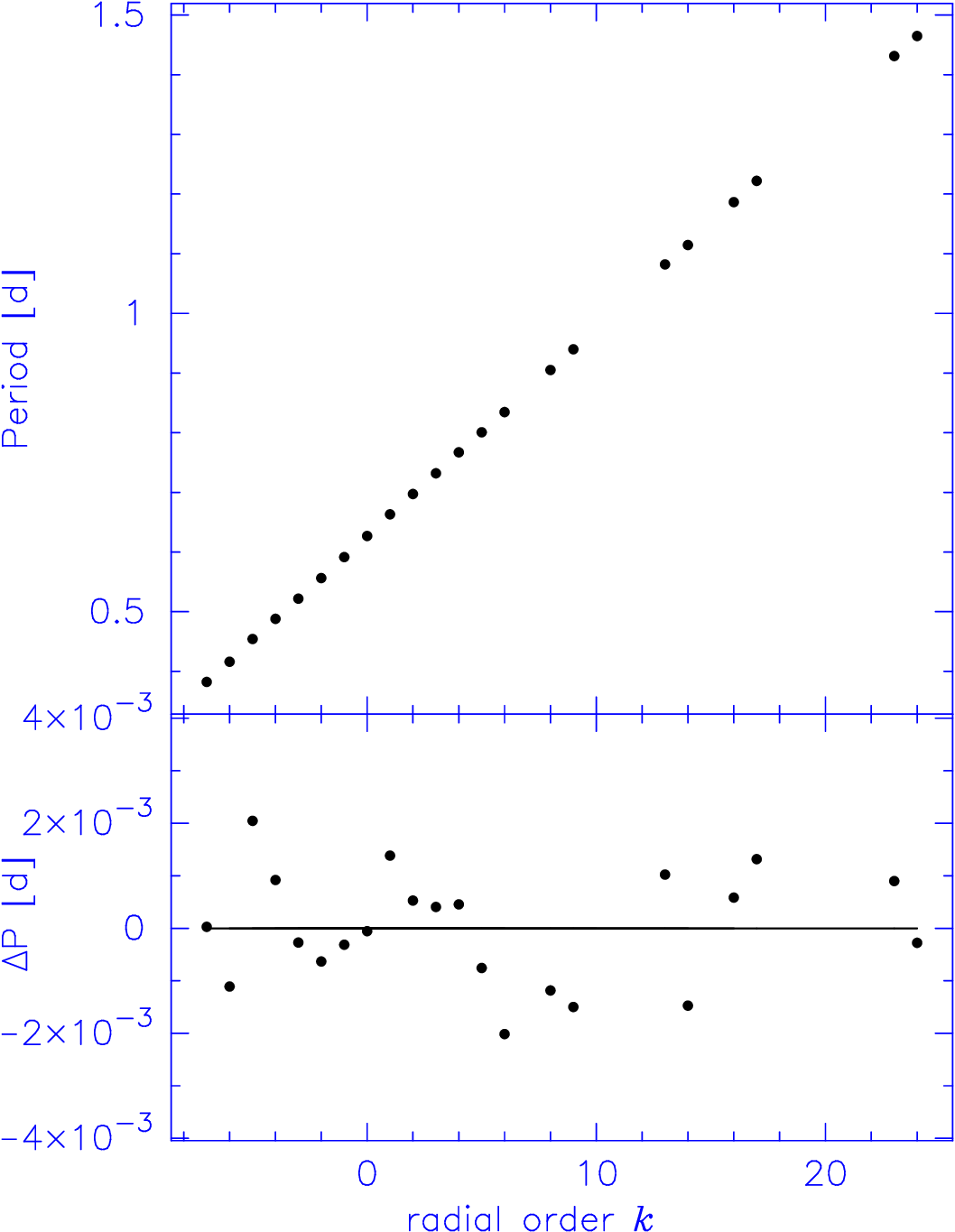}
\caption[]{\gd\ periods that obey the asymptotic period spacing law as a function of radial order $k$ with
an arbitrary zero point [top].
The bottom part shows the residuals of this period spacing; a straight line has been added for better visibility.
}
\label{fig06}
\end{figure}

We searched for equidistant \gd\ periods: a series of 22 asymptotic periods is found, with a 
mean separation $\Delta P=0.03493 \pm 0.00007$\,d 
(Table\,4, Fig.\,\ref{fig06}). 
\begin{table}
\begin{center}
\caption[]{List of the asymptotic frequencies.} 
\begin{tabular}{lrrr}
\hline
\multicolumn{1}{c}{Frequency} &
\multicolumn{1}{c}{Period} &
\multicolumn{1}{c}{Amplitude} &
\multicolumn{1}{c}{$k$} \\
\hline
$F_{109}$\tablefootmark{*} &   0.3824 &   0.16 & -7 \\
$F_{120}$\tablefootmark{*} &   0.4161 &   0.14 & -6 \\
$F_{17}$                   &   0.4542 &   0.59 & -5 \\
$F_{60}$\tablefootmark{*}  &   0.4880 &   0.26 & -4 \\
$F_{11}$                   &   0.5218 &   2.28 & -3 \\
$F_{6}$                    &   0.5564 &   4.21 & -2 \\
$F_{5}$                    &   0.5916 &   4.42 & -1 \\
$F_{2}$                    &   0.6268 &   8.45 &  0 \\
$F_{124}$                  &   0.6632 &   0.15 &  1 \\
$F_{3}$                    &   0.6973 &   6.24 &  2 \\
$F_{4}$                    &   0.7321 &   5.55 &  3 \\
$F_{113}$                  &   0.7670 &   0.16 &  4 \\
$F_{7}$                    &   0.8008 &   3.99 &  5 \\
$F_{8}$                    &   0.8344 &   3.52 &  6 \\
$F_{67}$                   &   0.9051 &   0.23 &  8 \\
$F_{135}$                  &   0.9398 &   0.13 &  9 \\
$F_{71}$                   &   1.0820 &   0.22 & 13 \\
$F_{15}$                   &   1.1145 &   0.74 & 14 \\
$F_{51}$\tablefootmark{*}  &   1.1864 &   0.28 & 16 \\
$F_{86}$                   &   1.2220 &   0.18 & 17 \\
$F_{66}$                   &   1.4312 &   0.22 & 23 \\
$F_{75}$                   &   1.4650 &   0.21 & 24 \\
\hline
\end{tabular}
\tablefoot{
Columns indicate the label of the frequency,
the period [d], the amplitude [mmag], and the order $k$ of the pulsation mode (arbitrarily 
shifted to $k=0$ for the main frequency $f_1$).
\tablefoottext{*}{These asymptotic frequencies are blended with combination frequencies.}}
\end{center}
\label{tab4}
\end{table}
In particular, the eight main \gd\ periods belong to these asymptotic series. 
The asymptotic periods extend from 0.382 to 1.465\,d, with a concentration of high-amplitude periods between 
0.525 and 0.834\,d.
These periods correspond to $g$-modes of the same radial degree $\ell$ with consecutive radial orders $k$. 
The measured $\Delta P$ separation is compatible with an $\ell=1$ $g$-modes \citep{bmm11}.
Four of these periods can be identified both as asymptotic and a combination between \gd\ modes. 
Longer observations would be necessary to separate the pikes that are blended in our frequency analysis.

Apart from the frequencies corresponding to linear combinations and those coupled with the orbital frequency 
(see Sect.\,6), 63 independent \gd\ frequencies remain.
These frequencies were sorted in terms of decreasing amplitudes and re-numbered starting from $f_1$. 

\section{The \ds\ domain}

\begin{figure}
\centering
\includegraphics[width=6.5cm,angle=-90.]{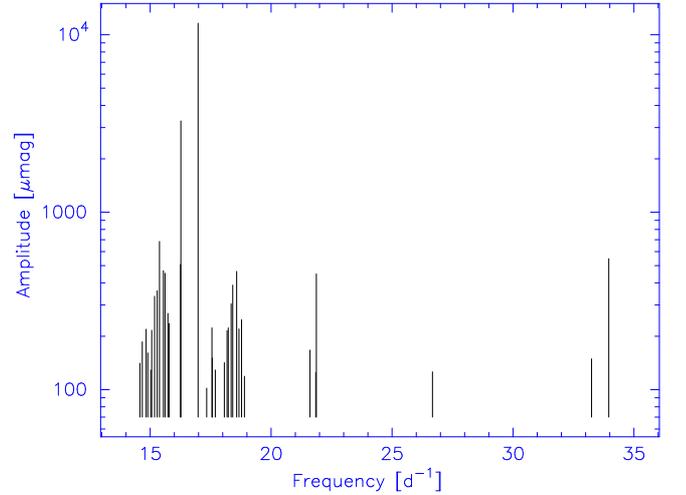}
\caption[]{Detected frequencies in the \ds\ domain.
The ordinates are on a logarithmic scale.
}
\label{fig07}
\end{figure}

\begin{figure}
\centering
\includegraphics[width=6.5cm,angle=-90.]{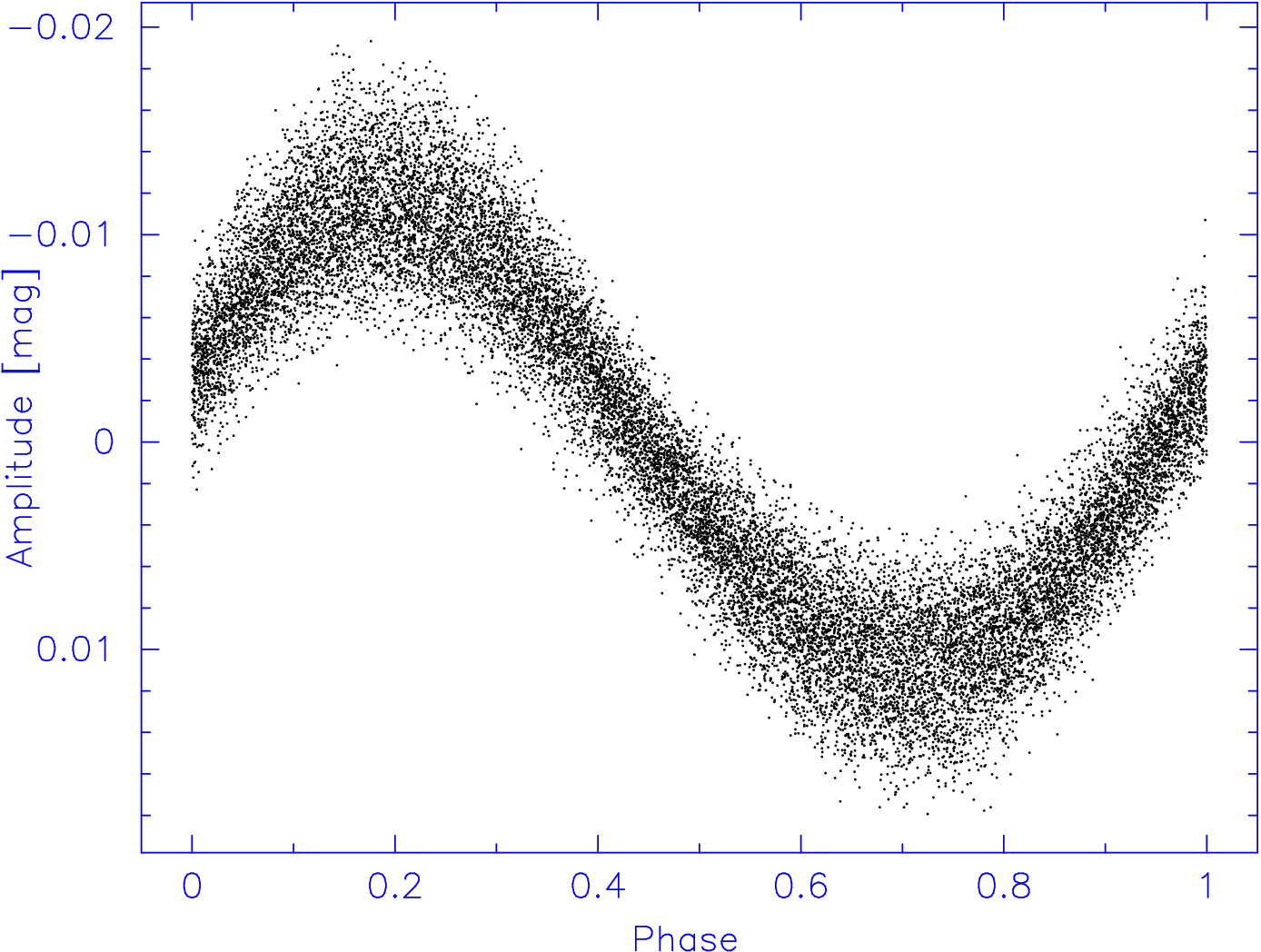}
\caption[]{Data phased with the main \ds\ frequency $F=16.9803$\,\cd.
}
\label{fig08}
\end{figure}

We retained 35 frequencies in the interval $[14.577;33.96]$\,\cd\ (Table\,2, Fig.\,\ref{fig07}). 
One frequency clearly dominates the spectrum, $F=16.9803$\,\cd,  with an amplitude $A=11.62$\,mmag. 
Figure\,\ref{fig08} represents a light curve phased with this frequency. 
The variation is quite sinusoidal, and only one low-amplitude harmonic ($2F=33.9604$\,\cd, $A=0.55$\,mmag) is found. 

The strong interactions between $g$- and $p$-modes revealed in 033 (Paper\,I)
are also present in 999.
Indeed, the first eight \gd\ frequencies $f_i$ are present in the \ds\ domain as $F \pm f_i$ (Table\,5).
\begin{table}
\begin{center}
\caption[]{List of the coupled \gd\ and \ds\ frequencies.}
\begin{tabular}{lcrr}
\hline
\multicolumn{1}{c}{Frequency} &
\multicolumn{1}{c}{Coupling} &
\multicolumn{1}{c}{$\Delta$} &
\multicolumn{1}{c}{Ratio} \\
\hline
$F_{16}$  & $F-f_1$ &  2 & 12.4 \\
$F_{23}$  & $F+f_1$ &  1 & 18.2 \\
$F_{22}$  & $F-f_2$ &  1 & 13.3 \\
$F_{32}$  & $F+f_2$ &  2 & 16.1 \\
$F_{24}$  & $F-f_3$ &  3 & 12.2 \\
$F_{44}$  & $F+f_3$ &  2 & 18.2 \\
$F_{34}$  & $F-f_4$ &  5 & 12.2 \\
$F_{74}$  & $F+f_4$ &  1 & 20.1 \\
$F_{38}$  & $F-f_5$ &  3 & 12.5 \\
$F_{56}$  & $F+f_5$ &  3 & 17.0 \\
$F_{52}$  & $F-f_6$ &  1 & 14.9 \\
$F_{69}$  & $F+f_6$ &  1 & 17.9 \\
$F_{62}$  & $F-f_7$ &  2 & 14.9 \\
$F_{70}$  & $F+f_7$ & 10 & 16.3 \\
$F_{76}$  & $F-f_8$ & 14 & 10.5 \\
$F_{147}$ & $F+f_8$ &  2 & 19.1 \\
\hline
\end{tabular}
\tablefoot{
Columns are the frequency label, the related coupling,
$\Delta=|F \pm f-F_{\rm measured}|$ in units of $10^{-4}$\,\cd, and the amplitude ratio
$A_{f_i}/(A_{F \pm f_i})$.}
\end{center}
\label{tab5}
\end{table}
The mean amplitude ratio $A_{f_i}/A_{F \pm f_i}$ is equal to $15.3 \pm 1.1$ with a lower value for the 
$A_{f_i}/A_{F+fi}$ ratio ($17.8 \pm 1.6$) than for the $A_{f_i}/A_{F-f_i}$ one ($12.8 \pm 1.5$). 
For 033 the ratio was equal to 4.1; in this latter star we also detected 
interaction with the $2F$ and $3F$ harmonics with ratios of 15 and 53, respectively. 

Getting rid of the coupled frequencies, only eight really independent \ds\ frequencies remain in the range $[16.253; 26.669]$\,\cd\ (Table\,6). 
\begin{table}
\begin{center}
\caption[]{List of the independant \ds\ frequencies.}
\begin{tabular}{lcrr}
\hline
\multicolumn{3}{c}{Frequency} &
\multicolumn{1}{c}{Amplitude} \\
\multicolumn{3}{c}{[\cd]} &
\multicolumn{1}{c}{[mmag]} \\
\hline
$F_{1}$   & $F$   & 16.9803 & 11.623 \\
$F_{21}$  & $p_1$ & 16.2530 &  0.508 \\
$F_{25}$  & $p_2$ & 21.8711 &  0.449 \\
$F_{77}$  & $p_3$ & 17.5521 &  0.223 \\
$F_{101}$ & $p_4$ & 21.6053 &  0.167 \\
$F_{116}$ & $p_5$ & 17.5674 &  0.151 \\
$F_{145}$ & $p_6$ & 26.6690 &  0.126 \\
$F_{144}$ & $p_7$ & 21.8441 &  0.125 \\
\hline
\end{tabular}
\end{center}
\label{tab6}
\end{table}

The amplitude of the main frequency $F=16.9303$\,\cd\ is almost 20 times larger than the amplitude 
of the other \ds\ frequencies. 
The period ratio $p_2/F=0.776$ is equal to the well-known period ratios of the fundamental mode to the first overtone $P_1/P_0=0.772$
(\citet{f81}, \citet{psn05}). 
If $F$ is really the radial fundamental mode, the period-luminosity relation of \citet{tbd02}, the period associated with
$F$ (0.05889\,d) corresponds to an absolute magnitude $M_v=2.4$\,mag and an A8\,V spectral type for the pulsating star. 
These values are consistent with those derived from the orbital parameters (Sect.\,3).
However, spectroscopic and multicolor observations are essential to confirm or invalidate the mode identification of $F$ and $p_2$.

\section{Binarity-Pulsation relationship}

Beside the different couplings detected between the pulsation frequencies, several couplings are detected with
the orbital frequency $f_{\rm orb}=0.35595$\,\cd.
These concern \ds\ as well as \gd\ frequencies.
The same couplings are detected in the period analysis using the entire data set prewhitened from the binary model. 
Therefore it cannot be an effect of aliasing due to the data gaps. 
The detected couplings are listed in Table\,7.

\begin{table}
\begin{center}
\caption[]{Linear combinations of \gd\ (first part) and \ds\ (second part) frequencies detected involving the 
orbital frequency $f_{\rm orb}$.}
\begin{tabular}{llrr}
\hline
\multicolumn{1}{c}{Frequency} &
\multicolumn{1}{c}{Combination} &
\multicolumn{1}{c}{$\Delta$} &
\multicolumn{1}{c}{Ratio} \\
\hline
$F_{39}$  & $f_1+2f_{\rm orb}$     &    6 & 26.4 \\
$F_{49}$  & $f_2+2f_{\rm orb}$     &    2 & 22.2 \\
$F_{93}$  & $f_3+2f_{\rm orb}$     &    9 & 29.1 \\
$F_{120}$ & $f_4+2f_{\rm orb}$     &    8 & 12.1 \\
$F_{97}$  & $f_6+2f_{\rm orb}$     &    9 & 22.3 \\
\hline
$F_{9}$   & $F-2f_{\rm orb}$       &    2 &  3.6 \\
$F_{141}$ & $F+2f_{\rm orb}$       &   12 & 90.0 \\
$F_{118}$ & $2F-2f_{\rm orb}$      &    8 &  3.7 \\
$F_{90}$  & $(F-f_1)-2f_{\rm orb}$ &    4 &  3.7 \\
$F_{80}$  & $(F-f_2)-2f_{\rm orb}$ &    6 &  2.1 \\
$F_{107}$ & $(F-f_3)-2f_{\rm orb}$ &    7 &  2.8 \\
$F_{133}$ & $(F-f_4)-2f_{\rm orb}$ &   11 &  2.6 \\
$F_{121}$ & $(F+f_5)+2f_{\rm orb}$ &    2 &  1.7 \\
$F_{134}$ & $(F-f_6)-2f_{\rm orb}$ &    4 &  2.1 \\
\hline
\end{tabular}
\tablefoot{
In addition to the frequency label and the linear combination, the differences 
$\Delta$ between the linear combination and the actually measured values [$\times 10^{-4}$\,\cd], 
as well as the ratio of the ``parent'' frequency amplitude to the amplitude of the combined frequency are provided.}
\end{center}
\label{tab7}
\end{table}

\begin{figure}
\centering
\includegraphics[width=6.5cm,angle=-90.]{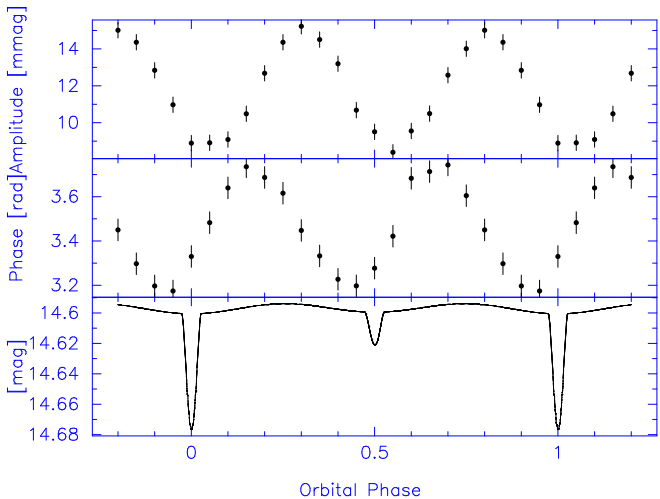}
\caption[]{Variation of the amplitude (top) and phase (middle) of the $F$ frequency as a function
of the orbital phase. For clarity, the bottom curve represents the fit of the light curve represented
in Fig.\,2. 
}
\label{fig09}
\end{figure}

The relative amplitudes of the $\pm 2f_{\rm orb}$ combinations are much larger for the high frequencies 
than for the low ones. 
The presence of sidelobes around all the main frequencies can be 
interpreted as amplitude variations during the orbital revolution. 
As the amplitude of $F-2f_{\rm orb}$ is particularly large ($A=3.27$\,mmag), we decided to measure the amplitude 
of the radial mode $F$ at different orbital phases. 
We sampled the orbital period in 20 bins, each containing about 1000 measurements, and analyzed separately each 
data set. 
We forced the frequency $F=16.9803$\,\cd\ and calculated the amplitudes and phases by least-square fits. 
The results are presented in Fig.\,\ref{fig09}. 
The amplitudes and phases present two perfect double-wave curves during the orbital period. 
The amplitudes varies over a range of $6.4$\,mmag, which is consistent with the amplitude of $F-2f_{\rm orb}$.
This variation is probably at the origin of the larger dispersion at the extrema in the $F$-phased light curve presented
in Fig.\,\ref{fig08}. 
The phase varies over a range of $0.076$ orbital period. 
The amplitude minima occur 0.05 orbital period (0.14\,d) after the two eclipses. 
We were not able to detect similar amplitude variations for the \gd\ frequencies, which is not surprising since 
the amplitudes of the $f_i + 2f_{\rm orb}$ frequencies are very low.

\section{Discussion}

After 033 (Paper\,I), the star 999 is the second one to present
a strong coupling between \gd\ and \ds\ modes.
In both stars, a large-amplitude \ds-mode is dominant, followed by 
relatively large-amplitude $g$-modes. 
The presence of the same sidelobes around the main \ds\ frequency in physically different 
stars (999 is cooler and rotates faster than 033) means that these
couplings are probably a general characteristic of hybrid \gd-\ds\ stars. 
We are convinced that the model of \citet{kjh93}, who predict such coupling in the Sun, is realistic. 
The low $g$-mode frequencies trapped in the stellar interior produce oscillatory perturbations of the
$p$-mode cavities and ``cause the formation of a pair of weak spectral
sidelobes symmetrically placed about the unperturbed p-mode frequency''.

The amplitude of the pulsation in 999 is smaller than that in 033: 
$A=11.62$\,mmag and $A=27.08$\,mmag for the main \ds\ mode, and
$A=8.45$\,mmag and $A=11.15$\,mmag for the main \gd\ mode. 
The larger ratio $A(f_i)/A(F \pm f_i)$ in 999 (15 compared to 4 for 033) is probably
related to the lower amplitude of the main \ds\ mode. 
But, as mentioned in Paper\,I, the relation between the $F$-amplitude and that ratio is not
linear. 
Detection of such couplings and the measurement of their ratios in other
hybrid stars will give interesting information about the internal structure
of these stars. 
Moreover, the eventual generalization of this phenomenon to hybrid $\beta$\,Cephei-SPB stars should be explored.

Another kind of relationship between $p$- and $g$-modes has been detected by \citet{bfb12} in the hybrid 
\gd-\ds\ star KIC\,8054146. 
Indeed, in this very fast rotator with low-amplitude modes, \citet{bfb12} did not detect any sidelobes 
around the main $p$-mode. 
On the other hand, they state ``the four dominant $g$-modes determine the spacing of the higher
frequencies in and beyond the \ds\ $p$-mode frequency domain''. 
These authors claimed that ``this unusual behaviour may be related to the very rapid
rotation of the star.''

Still in KIC\,8054146, \citet{bfb12} also detected strong amplitude variability with timescales of 
months and years for the four dominant low frequencies. 
To detect such a long-term modulation, we analyzed separately the first and second half of our data set.
No significant evolution is noted between these two sets.
Of course, the length of the CoRoT observations is much shorter than the Kepler ones, so we are not able to 
detect very long-term variations. 
However, if 999 had large variations similar to those of KIC\,8054146, we should have detected them.  
Since the value of the four main frequencies of KIC\,8054146 is comparable to the rotation frequency, 
these amplitude variations may be related to the very fast rotation.

The asymptotic series of periods separated by $\Delta P=0.03493$\,d
corresponds to a family of $\ell=1$ modes with successive $k$ radial orders. 
\citet{mmn08} computed the evolutionary effect on the variation of $\Delta P$. 
In three of their models (with a central hydrogen abundance equal to $X_C=0.1$, 0.3 and 0.5), 
the fluctuations are larger than 0.01\,d, whereas the dispersion is about 0.002\,d in our case. 
Therefore, only the model with a central hydrogen abundance $X_C=0.7$ is compatible with the 
stability of $\Delta P$ observed in 999.
After elimination of all the combination or coupled frequencies and
the asymptotic series, 40 independent \gd\ frequencies remain. 
Since only low $\ell$-modes can be detected in photometry, other asymptotic series certainly exist,
but we were not able to detect them.

Finally, the double-wave variation of the amplitude and the phase of the main \ds\ frequency gives rise to 
$F \pm 2 f_{\rm orb}$ combinations. 
Such combinations were also detected for five of the main \gd\ frequencies, but with much lower amplitudes. 
Because the $2 f_{\rm orb}$ frequency corresponds to the elliptic luminosity variations, this phenomenon could be 
related to the geometry of the star: the minimum of the amplitude occurs when the visible surface of the star 
is close to its minimum (but not at minimum). 
Therefore,  this latter occurence, together with the presence of a small but real phase lag between the eclipse 
and the double wave curve, points rather towards tidal effects. 

\begin{acknowledgements}
The authors are grateful to the anonymous referee for useful remarks and comments.
The authors also thank the L3 students Fanny Girard and Matthieu Grau.
\end{acknowledgements}

\end{document}